\documentclass[12pt, draftcls, onecolumn]{IEEEtran}
\usepackage{setspace}
\usepackage{amsmath,amsfonts}
\usepackage{algorithmic}
\usepackage{algorithm}
\usepackage{array}
\usepackage[caption=false,font=normalsize,labelfont=sf,textfont=sf]{subfig}
\usepackage{textcomp}
\usepackage{stfloats}
\usepackage{url}
\usepackage{color}
\usepackage{verbatim}
\usepackage{graphicx}
\usepackage{cite}
\begin{document}

\title{\Large A Tutorial on Holographic MIMO Communications—Part III: Open Opportunities and Challenges}
\author{Jiancheng An,~\IEEEmembership{Member,~IEEE}, Chau Yuen,~\IEEEmembership{Fellow,~IEEE},\\Chongwen Huang,~\IEEEmembership{Member,~IEEE}, M\'erouane Debbah,~\IEEEmembership{Fellow,~IEEE},\\H. Vincent Poor,~\IEEEmembership{Life Fellow,~IEEE}, and Lajos Hanzo,~\IEEEmembership{Life Fellow,~IEEE}\\
%\author{Jiancheng An,~\IEEEmembership{Member,~IEEE}, Chau Yuen,~\IEEEmembership{Fellow,~IEEE},\\
%Chongwen Huang,~\IEEEmembership{Member,~IEEE}, M\'erouane Debbah,~\IEEEmembership{Fellow,~IEEE},\\
%H. Vincent Poor,~\IEEEmembership{Life Fellow,~IEEE}, and Lajos Hanzo,~\IEEEmembership{Life Fellow,~IEEE}\\
\emph{(Invited Paper)}
\thanks{This research is supported by the Ministry of Education, Singapore, under its MOE Tier 2 (Award number MOE-T2EP50220-0019). This work was supported by the Science and Engineering Research Council of A*STAR (Agency for Science, Technology and Research) Singapore, under Grant No. M22L1b0110. The work of Prof. Huang was supported by the China National Key R\&D Program under Grant 2021YFA1000500, National Natural Science Foundation of China under Grant 62101492, Zhejiang Provincial Natural Science Foundation of China under Grant LR22F010002, Zhejiang University Global Partnership Fund, Zhejiang University Education Foundation Qizhen Scholar Foundation, and Fundamental Research Funds for the Central Universities under Grant 2021FZZX001-21. H. V. Poor would like to acknowledge the financial support of the U.S. National Science Foundation under Grant CNS-2128448. L. Hanzo would like to acknowledge the financial support of the Engineering and Physical Sciences Research Council projects EP/W016605/1 and EP/X01228X/1 as well as of the European Research Council's Advanced Fellow Grant QuantCom (Grant No. 789028). \emph{(Corresponding author: Chau Yuen.)}}
\thanks{J. An is with the Engineering Product Development Pillar, Singapore University of Technology and Design, Singapore 487372 (e-mail: jiancheng\_an@sutd.edu.sg). C. Yuen is with the School of Electrical and Electronics Engineering, Nanyang Technological University, Singapore 639798 (e-mail: chau.yuen@ntu.edu.sg). C. Huang is with College of Information Science and Electronic Engineering, Zhejiang-Singapore Innovation and AI Joint Research Lab and Zhejiang Provincial Key Laboratory of Info. Proc., Commun. \& Netw. (IPCAN), Zhejiang University, Hangzhou 310027, China. (e-mail: chongwenhuang@zju.edu.cn ). M. Debbah is with Khalifa University of Science and Technology, P O Box 127788, Abu Dhabi, UAE (email: merouane.debbah@ku.ac.ae). H. Vincent Poor is with the Department of Electrical and Computer Engineering, Princeton University, Princeton, NJ 08544 USA (e-mail: poor@princeton.edu). L. Hanzo is with the School of Electronics and Computer Science, University of Southampton, SO17 1BJ Southampton, U.K. (e-mail: lh@ecs.soton.ac.uk).}}
\maketitle

\begin{abstract}
Holographic multiple-input multiple-output (HMIMO) technology, which uses spatially continuous surfaces for signal transmission and reception, is envisioned to be a promising solution for improving the data rate and coverage of wireless networks. In Parts I and II of this three-part tutorial on HMIMO communications, we provided an overview of channel modeling and highlighted the state-of-the-art in holographic beamforming. In this part, we will discuss the unique properties of HMIMO systems, highlighting the open challenges and opportunities that arise as the transceiver array apertures become denser and electromagnetically larger. Additionally, we explore the interplay between HMIMO and other emerging technologies in next-generation networks.
\end{abstract}


\section{Introduction}
\IEEEPARstart{H}{olographic} multiple-input multiple-output (HMIMO) communications offer a groundbreaking approach to manipulate the electromagnetic (EM) field with unprecedented flexibility, hence holding tremendous promise in terms of enhancing next-generation wireless networks \cite{arXiv_2022_Gong_Holographic, TCOM_2022_An_Low}. The underlying concept of HMIMO schemes relies on spatially continuous transceiver apertures associated with an almost infinite number of antennas and minuscule element spacing. A practical technique of implementing HMIMO schemes relies on utilizing intelligent metasurfaces, which are created from advanced meta-materials and micro-electromechanical systems \cite{DSP_2019_Bjornson_Massive}. The metasurface technology provides a seamless interface between the digital, circuit, and EM domains, resulting in a completely new information theoretic paradigm relying on optimized HMIMO communications from a fundamental physical perspective \cite{Proc_2022_Rnezo_Communication}.

In Parts I and II of this three-part tutorial, we examined the most recent advances in channel modeling, channel estimation, performance analysis, and holographic beamforming within the context of HMIMO communications. The newly developed HMIMO channel model, which is based on EM propagation theory, provides a solid foundation both for developing sophisticated holographic beamforming schemes and for analyzing the HMIMO capacity. A significant amount of research has showcased the enormous potential of HMIMO in achieving the ultimate performance limits of wireless communications.

However, to attain these potential benefits, a substantial community effort is required for solving the associated technological challenges \cite{PC_2017_Black_Holographic}. A notable issue is the limited tuning capability of practical intelligent metasurfaces, which is constrained by the specific control circuit used. For example, in a waveguide-fed metasurface, the radiated signals leaked from the substrate may experience significant attenuation, especially, when the number of radiating elements escalates \cite{WC_2021_Shlezinger_Dynamic}. Furthermore, the close proximity of a massive number of radiating elements will result in the mutual coupling, leading to a reduction in radiation efficiency. At the time of writing, there is a distinct lack of real-world measurements and communication experiments that can validate the expected multiplexing gain and enhanced directivity under various hardware limitations.

In this letter, we provide a comprehensive overview of the key research challenges in HMIMO communications and their potential to inspire future research. The rest of this letter is organized as follows. Section \ref{sec2} sheds light on the potential research directions related to HMIMO communications, while Section \ref{sec3} examines the interplay between HMIMO and a range of other emerging technologies. Finally, Section \ref{sec4} concludes the letter and provides a general discussion of the three-part tutorial.
\section{Potential Research Directions}\label{sec2}
\subsection{Hardware Architecture}
Again, an efficient way of implementing HMIMO is through the use of cost- and energy-efficient metasurfaces, which can significantly reduce the number of radio frequency (RF) chains required for achieving comparable performance to the conventional full-digital MIMO architectures. For example, Shlezinger \emph{et al.} \cite{WC_2021_Shlezinger_Dynamic} designed a metasurface-based antenna to generate holographic waveforms via a waveguide-fed structure. While advanced metamaterial technology has made it possible to dynamically manipulate EM waves with fine granularity, there is a clear trade-off between flexibility in shaping the EM field, complexity, and energy consumption. Therefore, the holographic capabilities under practical hardware constraints deserve further investigation. In \cite{Access_2020_Dai_Reconfigurable}, \emph{Dai et al.} have manufactured a 256-element metasurface prototype, where positive intrinsic-negative (PIN) diodes were utilized to realize $2$-bit phase shifting. It was demonstrated that at a millimeter wave frequency of $28.5$ GHz, the metasurface attains an antenna gain of $19.1$ dBi \cite{Access_2020_Dai_Reconfigurable}.

Moreover, advanced HMIMO hardware designs tend to rely on a multi-layer architecture mimicking a neural network and they perform signal processing directly in the wave domain \cite{arxiv_2023_An_Stacked}. This has the potential to significantly reduce the processing latency, as wave-based processing can be realized at the speed of light. Specifically, An \emph{et al.} \cite{ICC_2023_An_Stacked} developed a bespoke deep neural network by stacking multiple layers of metasurfaces at the transceiver. Such a stacked intelligent metasurface (SIM) architecture is capable of implementing transmit precoding and receiver combining in the EM wave domain, hence bypassing the need for digital beamforming. Additionally, an SIM-aided transceiver can perform adequately even using low-resolution analog-to-digital converters and a moderate number of RF chains, which significantly reduces the hardware costs and energy consumption, while mitigating the transmit precoding (TPC) delay as a benefit of operating in the wave domain.
\subsection{EM Information Theory (EMIT)}
Additionally, it should be noted that the development of MIMO technology is founded on Shannon’s information theory, which, however, neglects the underlying physical phenomena of EM wave propagation. By combining Shannon's and Maxwell's theories, a new interdisciplinary framework called EMIT has emerged for evaluating the fundamental limits of wireless communications \cite{arXiv_2022_Gong_Holographic}. EMIT treats wireless communications as a functional analysis problem and concentrates on identifying the optimal EM functions at the transceiver ends. As such, the intrinsic capacity of the continuous space can be characterized independently of the specific transmission technology, of the number of antenna elements, and of their layout.

To achieve this ambitious goal, Li \emph{et al.} \cite{TAP_2023_Li_An} have developed an efficient EMIT-based HMIMO communication model, where the EM coupling operator in free space is characterized by a pair of key parameters: the EM effective capacity and the path loss. Based on dyadic Green’s function and matrix mode analysis, the EM effective capacity is given by \cite{TAP_2023_Li_An}
\begin{align}\label{eq1}
 C_{\textrm{eff}}=\prod_{i=1}^{\min\left ( N_{t},N_{r} \right )}e^{-\sigma _{i}'\ln\left ( \sigma _{i}' \right )},
\end{align}
where $\sigma _{i}'=\sigma _{i}/\left ( \sum_{i}\sigma _{i} \right )$ represents the normalized singular values of the coupling operator, while $N_t$ and $N_r$ denote the number of antennas at the transmitter and receiver, respectively. Furthermore, a novel group-T-matrix-based multiple scattering fast algorithm was developed for describing a representative inhomogeneous EM space. Based on practical measurements of a cylindrical array in a microwave anechoic chamber, it is demonstrated that the EMIT-based model is more efficient and reliable than full-wave simulation, capable of analyzing the EM channel within 10\% of the time. This makes it suitable for cluster models having arbitrary distribution, size, and material, providing an efficient and reliable method for guiding the transmit power allocation in complex scattering environments and offering new insights into the extraction of suitable parameters using computational electromagnetism.

To summarize, the connection between classic information theory and EM wave transmission theory is capable of unleashing the full potential of HMIMO systems in the spatial domain and of guiding their practical deployment.
\subsection{EM Field Sampling}
The effective sampling and reconstruction of EM waves are crucial in HMIMO communications for the associated signal processing. To this end, it is necessary to develop the associated spatial domain sampling theory. Similar to the time-frequency sampling, a specific minimum number of samples has to be gleaned from the spatially continuous surface for accurately reconstructing the EM field without substantial information loss. When considering classical half-wavelength rectangular sampling, the Nyquist density (in samples/m$^{2}$) is determined by $\mu _{\mathcal{R}} = 4/\lambda ^{2}$ with $\lambda$ representing the wavelength. Note that half-wavelength sampling involves a portion of evanescent waves that occur only when the receiver is located close to the scatterers, which leads to an under-utilization of the wavenumber spectrum in practical scenarios.

By leveraging the multidimensional sampling theorem and Fourier theory, Pizzo \emph{et al.} \cite{TSP_2022_Pizzo_Nyquist} studied the Nyquist sampling and reconstruction of an EM field under arbitrary scattering conditions. Specifically, the three-dimensional (3D) wireless propagation is modeled as a two-dimensional (2D) space-invariant system having low-pass filtering behavior. The spatial bandwidth is determined by the selectivity of the underlying scattering environment and it is maximized under isotropic propagation. To elaborate on the associated Nyquist sampling, next we consider both isotropic and non-isotropic scattering scenarios.
\begin{figure}[!t]
\centering
\includegraphics[width=12cm]{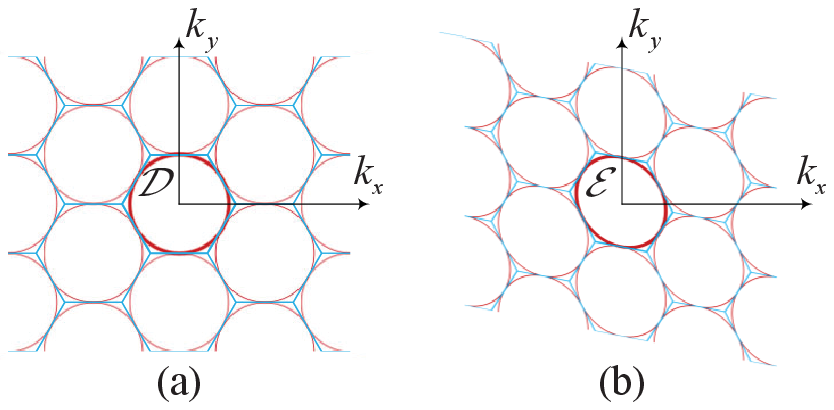}
\caption{(a) Nyquist sampling under isotropic scattering: circle packing in the wavenumber domain; (b) Nyquist sampling under non-isotropic scattering: ellipse packing in the wavenumber domain.}
\label{fig_1}
\end{figure}
\subsubsection{Isotropic Scattering Environment}
In an isotropic scattering environment, the Nyquist periodicity matrix can be geometrically found by solving a packing problem of circles formulated as
\begin{align}
 \mathcal{D}=\left \{ \mathbf{k}\in \mathbb{R}^{2}|\left \| \mathbf{k} \right \|\leq \kappa \right \},
\end{align}
with equal radii of $\kappa =2\pi /\lambda $. The solution is illustrated in Fig. \ref{fig_1}(a), which involves \emph{i)} inscribing each circle with a regular hexagon of side lengths $2\kappa /\sqrt{3}$ and \emph{ii)} replicating this perfectly tiling hexagonal arrangement periodically to avoid aliasing. This results in a hexagonal sampling with a Nyquist density of $\mu _{\mathcal{D}}=2\sqrt{3}/\lambda ^{2}$. Compared to rectangular sampling, the hexagonal sampling has a sampling efficiency gain of 
\begin{align}
 1-\mu _{\mathcal{D}}/\mu _{\mathcal{R}} = 1-\sqrt{3}/2=13.4\%.
\end{align}
This means that the hexagonal sampling requires 13.4\% fewer samples than rectangular sampling.
\subsubsection{Non-Isotropic Scattering Environment}
Nyquist sampling in a non-isotropic scattering environment is exceedingly complex due to the irregular shape of the support $\mathcal{K}$. To address this issue, a suboptimal strategy involves using a larger and connected support in the wavenumber domain. For analytical tractability, we opt for a 2D ellipse embedding of $\mathcal{E}\supseteq \mathcal{K}$, which is defined by \cite{TSP_2022_Pizzo_Nyquist}
\begin{align}
 \mathcal{E}=\left \{ \mathbf{k}\in \mathbb{R}^{2}|\mathbf{k}^{T}\mathbf{G}^{-1}\mathbf{k}\leq \kappa ^{2} \right \},
\end{align}
where $\mathbf{G}\in \mathbb{R}^{2\times 2}$ is a symmetric positive definite matrix that specifies the shape factors of the ellipse. As a result, the Nyquist periodicity matrix for the elliptical embedding of $\mathcal{K}$ is geometrically formulated as an ellipse packing problem. Consequently, the sampling scheme in this case is a rotated and scaled version of the one obtained under isotropic scattering conditions, i.e., an elongated hexagonal sampling, as shown in Fig. \ref{fig_1}(b). The stretching of hexagonal sampling is uniquely determined by the angular selectivity of the scattering. The Nyquist density is computed as $\mu _{\mathcal{E}}=2\sqrt{3d_{1}d_{2}}/\lambda ^{2}$, where $\sqrt{d_{1}}$ and $\sqrt{d_{2}}$ are the normalized lengths of the semi-axes of $\mathcal{E}$ \cite{TSP_2022_Pizzo_Nyquist}. Hence, the elongated hexagonal sampling has a sampling efficiency gain of
\begin{align}\label{eq3}
 1-\mu _{\mathcal{E}}/\mu _{\mathcal{R}}=1-\sqrt{3d_{1}d_{2}}/2,
\end{align}
compared to half-wavelength sampling. For example, if the scatterers occupy a solid angle spanning the entire azimuthal horizon and half range in elevation angle, we have $\sqrt{d_{1}}=1$ and $\sqrt{d_{2}}=0.5$. In this case, the sampling efficiency gain in \eqref{eq3} is approximately equal to $38.8\%$. Improved sample reduction is expected as the scattering becomes more selective in the wavenumber domain. The specific Nyquist sampling matrix and interpolating function may be found in \cite{TSP_2022_Pizzo_Nyquist}.
\subsection{Mutual Coupling}
Mutual coupling is an inherent characteristic of transceiver surfaces that have closely packed radiating elements, and it affects both the radiation pattern and the impedance of the antenna element in practice. In conventional MIMO design, the antennas are usually placed at least half of a wavelength apart for minimizing mutual coupling and simplifying the system design. This results in a maximum array gain that is proportional to the number of antennas. However, recent research has indicated that by appropriately exploiting the mutual coupling of subwavelength-spaced antennas, it is possible to achieve super-directivity \cite{ACSSC_2019_Marzetta_Super}. This breakthrough could lead to significantly increased antenna array gains, thereby expanding the coverage area without increasing the transmit power. For HMIMO systems, the array gain is expected to reach the square of the number of radiating elements.

\begin{figure}[!t]
\centering
\includegraphics[width=8cm]{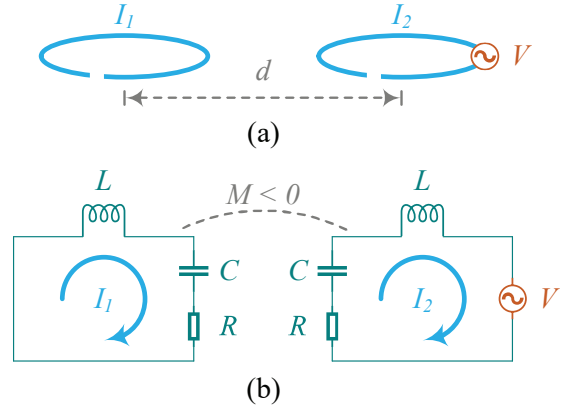}
\caption{(a) Schematic representation of a dimer of coupled ring resonators with only one element being driven by the external voltage, where $d$ is the distance between two vertical-oriented magnetic dipole radiators, $V$ is the external voltage, $I_1$ and $I_2$ are the current distribution in two elements. (b) The equivalent circuit, where $M$ represents mutual inductance, $L$, $C$, and $R$ are the self-inductance, the capacitance, and the resistance, respectively, of each meta-atom.}
\label{fig_2}
\end{figure}
Specifically, Williams \emph{et. al.} \cite{ICC_2020_Williams_A} studied large intelligent surface (LIS)-based HMIMO downlink transmission in a line-of-sight (LoS) environment. They introduced a communication model that takes into account the mutual coupling between antennas and demonstrated the potential benefits of the super-directivity attained by LISs. The directivity gain is theoretically unbounded as the element density of the surface increases for a constant aperture dimension. However, there are several practical challenges in the way of achieving those superior gains. These adverse factors include high ohmic losses due to the high currents involved, the need for accurately adjusting the excitation current, and scan blindness. More recently, Han \emph{et al.} \cite{ICC_2022_Han_Coupling} derived a beamformer for achieving super-directivity relying on a compact array and harnessing antenna coupling. The coupling matrix was derived using the spherical wave expansion method. The directivity of the printed dipole antenna array, which operated at $845$ MHz, was validated through full-wave EM simulations. In practice, the strong mutual coupling and extremely high excitation currents also result in a reactive field in the vicinity of the array, thus leading to significant losses for any internal resistance in the antennas, which reduces the efficiency of the array and makes it sensitive even to small random variations in the excitation. Using superconducting antennas might help alleviate the effects of ohmic loss.

Moreover, implementing super-directivity by relying on metamaterial elements may not require separately exciting each individual element. Specifically, Yan \emph{et al.} \cite{OJAP_2020_Yan_A} proposed an efficient technique for optimizing the directivity in metamaterial-inspired end-fire antenna arrays having strong inter-element coupling. As shown in Fig. \ref{fig_2}, dimers of magnetically coupled split-ring resonators were used, with only a single element being driven by an external source. They established the conditions of achieving super-directive current distributions, which impose constraints on the quality of the resonators and on their coupling with respect to the array size and the operating frequency. Nevertheless, this implementation shares the common impediments of all super-directive antennas, namely their limited bandwidth, high tolerance sensitivity, and low efficiency.
\subsection{Orbital Angular Momentum (OAM)}
\begin{figure}[!t]
\centering
\includegraphics[width=10cm]{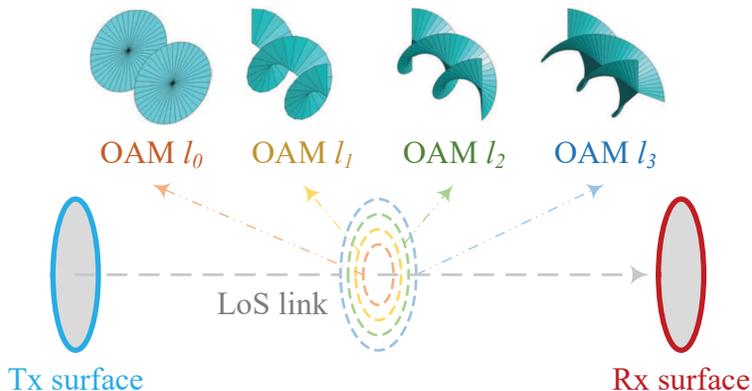}
\caption{Stylized model of an OAM-aided wireless communication system using four different OAM modes. The OAM modes are $L = 0$, $1$, $2$, and $3$, respectively.}
\label{fig_3}
\end{figure}

Increasing the antenna aperture's element density in HMIMO systems also facilitates the utilization of OAM multiplexing. Specifically, OAM is a unique component of the angular momentum of EM waves, which is generated when the electric and magnetic vector fields within a vortex have a constant phase, resulting in a helix-like shape that runs in the same direction as the wave's propagation. The vortex is characterized by a topological charge, which represents the number of twists that the light undergoes in a single wavelength, as shown in Fig. \ref{fig_3}. A fascinating aspect of OAM is that the momentum it carries can theoretically have an infinite number of eigenstates. As such, fully exploiting the OAM dimensions for information multiplexing could significantly improve the channel capacity.

In order to capitalize on the attainable multiplexing gain, Wu \emph{et al.} \cite{TAP_2021_Wu_Millimeter} designed a flat holographic lens antenna that is capable of simultaneously transmitting a pair of independent OAM beams in the millimeter wave (mmW) band. The design consists of a wideband C-shaped polarization-conversation transmit array element and of a broadband magneto-electric (ME) dipole antenna, which serve as its building block and feed source, respectively. The flat lens transforms the spherical wavefront from the pair of ME dipole sources into transmitted vortex waves having different topological charges, without the need for complicated OAM beam feed sources. By adjusting the geometric relationship between the feed sources and the flat lens, the residual OAM mode is effectively eliminated. The design achieves a high measured isolation of over $20$ dB for the twin-mode OAM channels, providing reliable transmission links simultaneously. Recently, Torcolacci \emph{et al.} \cite{GLOBECOM_2022_Torcolacci_OAM} explored the potential of exploiting the OAM property of the EM waves relying on LISs in near-field LoS propagation conditions for HMIMO communications. They proposed a set of simple OAM-related basis functions at the transmitter, based on the communication mode description of beams. To enhance the performance of the OAM-based transmission with the aid of LISs, they proposed a practical beam-focusing technique. Their numerical study demonstrated the feasibility of constructing several orthogonal channels, particularly when beam-focusing was harnessed for limiting the beams’ divergence.

In a nutshell, the powerful EM wave manipulation capabilities of HMIMO schemes make it possible to generate a much wider variety of beam modes than conventional MIMO systems. This breakthrough has the potential of revolutionizing wireless communications, paving the way for a paradigm shift from conventional eigenmode multiplexing to radical OAM mode multiplexing, resulting in a significant increase in system capacity \cite{TWC_2017_Ren_Line}.

\section{Interplay with Emerging Technologies}\label{sec3}
In this section, we will examine the potential benefits of integrating HMIMO schemes with emerging techniques in next-generation networks. Additionally, we will identify open challenges for further investigations.
\subsection{Wireless Power Transfer}
Wireless power transfer (WPT) offers a promising solution for charging wireless devices without relying on a wired infrastructure. This technology is particularly beneficial for charging devices having limited battery capacity. To enhance the efficiency and extend the operational range of WPT systems, metasurfaces can be harnessed for holographic beamforming. In particular, metamaterial-based rectennas can significantly improve the conversion efficiency from RF to direct current energy \cite{Proc_2022_Zhou_Metamaterials}. Metamaterials can also serve as either parasitic elements or loading components for improving wireless energy harvesting performance. By having metasurfaces, the reception becomes less sensitive to incident wave angle and polarization, leading to improvements in circuit size, beamwidth, and conversion efficiency. In general, there are three main ways that metasurface-based HMIMO schemes can improve the WPT performance: \emph{i) reflecting} the EM field in other directions to suppress the unintended EM radiation; \emph{ii) concentrating} the EM field between a transmitter and a receiver for extending the WPT distance; and \emph{iii) directing} the EM field to a receiver with dynamic control. As a benefit of extremely large antenna apertures, future RF-based WPT devices are expected to operate in the radiating near-field region, which will require both energy beam-focusing and radiated waveform design for implementing near-field radiation based WPT.
\subsection{Satellite Communications}
Ultra-dense low-Earth-orbit (LEO) satellite communication networks have the potential of providing high-rate data services \cite{JSAC_2021_Cao_Reconfigurable}. To compensate for the severe path loss of satellite communications, a spatially continuous aperture can achieve a high directional gain with a small antenna size. This makes HMIMO technology an excellent candidate for assisting satellite communications and for overcoming challenges such as power constraints and hardware limitations. In \cite{JSAC_2022_Deng_Holographic}, Deng \emph{et al.} investigated the employment of reconfigurable holographic surfaces (RHS) in support of multiple LEO satellites. For supporting seamless communication with multiple satellites, a satellite tracking scheme was proposed based on the temporal variation law of the satellite positions, which was capable of coping with the mobility of LEO satellites. A holographic beamforming algorithm was then developed for maximizing the sum rate. Their simulation results demonstrated that the RHS outperforms conventional phased arrays of the same size both in terms of its sum rate and its cost efficiency, thanks to the dense element-spacing of the RHS, as a benefit of allowing for more RHS elements. Overall, the use of the HMIMO technique in satellite communications significantly improves the quality of service (QoS).
\subsection{Integrated Sensing and Communication}
Integrating sensing and communication (ISAC) functionalities into a common hardware platform has substantial benefits \cite{TWC_2023_An_Fundamental}. It is expected that HMIMO has great potential in enhancing both the sensing and communication performance by providing a fine-grained high-gain beam pattern. For example, Zhang \emph{et al.} \cite{JSAC_2022_Zhang_Holographic} proposed a holographic ISAC scheme relying on metamaterial antennas, in which holographic beamforming is performed by an RHS of the base station (BS). By optimizing holographic beamforming, a desired beampattern can be generated in the direction of the targets with maximum gain, while meeting the QoS requirements of all communication users. The theoretical analysis demonstrated that densely packed radiating elements may increase the beamforming gain by at least 50\% compared to a conventional MIMO array of the same size, while also reducing the hardware costs. Beyond wireless sensing, next-generation wireless systems are expected to play a crucial role in wireless localization and positioning. The approximately continuous aperture of HMIMO schemes is capable of achieving a high localization resolution. To analyze the fundamental limits of electromagnetically large antenna arrays for localization, D'Amico \emph{et al.} \cite{TSP_2022_DAmico_Cramer} computed the Cram\'er-Rao Bound (CRB) for holographic positioning based on the EM wave propagation theory, which inherently depends on the source orientation. The maximum likelihood (ML) estimation based on the complete EM model achieves high estimation accuracy compared to conventional benchmarks.

In addition to the aforementioned extensions, the emergence of HMIMO schemes has sparked off intense research efforts by introducing large apertures having dense elements to various fields. For example, by incorporating HMIMO schemes into THz communications, one can potentially create a communication system having compelling benefits, such as simplified transceiver hardware architecture, high data rates, low latency, and improved transmission distance and coverage range. However, the combination of HMIMO schemes with these promising technologies is yet to be fully explored. As such, both the theoretical analysis and the practical algorithms of HMIMO communications are unexplored at the time of writing and open up new research opportunities.
\section{Conclusions}\label{sec4}
In this letter, we have presented a comprehensive overview of the major research challenges and recent progress in HMIMO communications. Specifically, we have discussed the issues of EMIT, EM-field sampling, mutual coupling, and potential OAM multiplexing. Furthermore, we have elucidated the key technical challenges and open research directions. Additionally, we examined the interplay of HMIMO schemes with other emerging technologies. While the HMIMO communication is still in its nascent stage, this three-part tutorial aims to inspire its future evolution and provide a convenient reference for those interested in this rapidly evolving field.

\bibliographystyle{IEEEtran}
\bibliography{ref}
\end{document}